\documentclass[
final,1p,times
]{elsarticle}

\makeatletter
\def\ps@pprintTitle{%
  \let\@oddhead\@empty
  \let\@evenhead\@empty
  \def\@oddfoot{\reset@font\hfil\thepage\hfil}
  \let\@evenfoot\@oddfoot
}
\makeatother

\usepackage[utf8]{inputenc}
\usepackage[T1]{fontenc}
\usepackage{xcolor}
\usepackage{listings}
\usepackage{graphicx}
\usepackage{bm}
\usepackage[normalem]{ulem}
\usepackage{url}
\usepackage{caption}
\usepackage{subcaption}

\definecolor{codegreen}{rgb}{0,0.6,0}
\definecolor{codegray}{rgb}{0.5,0.5,0.5}
\definecolor{codepurple}{rgb}{0.58,0,0.82}
\definecolor{backcolour}{rgb}{0.95,0.95,0.92}

\lstdefinestyle{mystyle}{
    backgroundcolor=\color{backcolour},
    commentstyle=\color{codegreen},
    keywordstyle=\color{magenta},
    numberstyle=\tiny\color{codegray},
    stringstyle=\color{codepurple},
    basicstyle=\ttfamily\footnotesize,
    breakatwhitespace=false,
    breaklines=true,
    captionpos=b,
    keepspaces=true,
    numbers=left,
    numbersep=5pt,
    showspaces=false,
    showstringspaces=false,
    showtabs=false,
    tabsize=2
}

\lstdefinelanguage{JavaScript}{
  keywords={typeof, new, true, false, catch, function, return, null, catch, switch, var, if, in, while, do, else, case, break},
  keywordstyle=\color{blue}\bfseries,
  ndkeywords={class, export, boolean, throw, implements, import, this},
  ndkeywordstyle=\color{darkgray}\bfseries,
  identifierstyle=\color{black},
  sensitive=false,
  comment=[l]{//},
  morecomment=[s]{/*}{*/},
  commentstyle=\color{purple}\ttfamily,
  stringstyle=\color{red}\ttfamily,
  morestring=[b]',
  morestring=[b]"
}

\graphicspath{{./}{./Figures/}}
\usepackage[hidelinks]{hyperref}

\begin{document}

\begin{frontmatter}

\title{Jupyter widgets and extensions for education and research in computational physics and chemistry}

\author[CECAM,epfl,marvel]{Dou Du\corref{contrib}}
\author[CECAM,marvel]{Taylor J. Baird\corref{contrib}}
\author[epfl,marvel]{Kristjan Eimre}
\author[CECAM,marvel]{Sara Bonella}
\author[psi,epfl,marvel]{Giovanni Pizzi\corref{mycorrespondingauthor}}
\cortext[mycorrespondingauthor]{Corresponding author}
\cortext[contrib]{Authors contributed equally}
\ead{giovanni.pizzi@psi.ch}

\address[CECAM]{CECAM Centre Européen de Calcul Atomique et Moléculaire, École Polytechnique Fédérale de Lausanne, CH-1015 Lausanne, Switzerland}

\address[epfl]{Theory and Simulation of Materials (THEOS), École Polytechnique Fédérale de Lausanne, CH-1015 Lausanne, Switzerland}

\address[marvel]{National Centre for Computational Design and Discovery
of Novel Materials (MARVEL), École Polytechnique Fédérale de Lausanne, CH-1015 Lausanne, Switzerland}

\address[psi]{Laboratory for Materials Simulations (LMS), Paul Scherrer Institut (PSI), CH-5232 Villigen PSI, Switzerland}

\begin{abstract}
Interactive notebooks are a precious tool for creating graphical user interfaces and teaching
materials. Python and Jupyter are becoming increasingly popular in this context, with Jupyter widgets at the core of the interactive functionalities. However, while 
packages and libraries which offer a broad range of general-purpose widgets exist, there is limited development of specialized widgets for computational physics, chemistry and materials science. This deficiency implies significant time investments for the development of effective Jupyter notebooks for research and education in these domains. 
Here, we present custom Jupyter widgets that we have developed to target the needs of these communities. These widgets constitute high-quality interactive graphical components and can be employed, for example, to visualize and manipulate data, or to explore different visual representations of concepts, clarifying the relationships existing between them. In addition, we discuss with one example how similar functionality can be exposed in the form of JupyterLab extensions, modifying the JupyterLab interface for an enhanced user experience when working with applications within the targeted scientific domains.
\end{abstract}


\end{frontmatter}

\begin{small}
\noindent
{\em Program Title:  Development of Jupyter widgets and extensions for computational physics and chemistry}  \\
{\em Developer's repository links}: \\
\url{https://github.com/osscar-org/widget-bzvisualizer}  \\
\url{https://github.com/osscar-org/widget-periodictable}  \\
\url{https://github.com/osscar-org/widget-bandsplot}  \\
\url{https://github.com/osscar-org/jupyterlab-mol-visualizer}  \\
{\em Licensing provisions:} MIT  \\
{\em Programming language: JavaScript, TypeScript, Python}                                   \\
{\em Nature of problem(approx. 50-250 words):}\\
 \\
 Interactive widgets can greatly enhance the learning process of students when used as components of educational notebooks, and streamline day-to-day research tasks of scientists when employed within research workflows leveraging the Jupyter ecosystem. Unfortunately, there is still a lack of specialized widgets tailored for use in computational physics or chemistry. Due to this, user experience and time-to-development or adoption of software tools of this kind are still poor. 

{\em Solution method(approx. 50-250 words):}\\

We have developed a number of high-quality and versatile Jupyter widgets and extensions, delivering customized visual software components for computational physics and chemistry. These tools considerably enhance education and research applications in these domains. Our widgets have already been adopted in several educational or research tools. The software is available and open source, facilitating adoption. 
\end{small}

\section{\label{sec:level1} Introduction}
In the past decade, the use of digital tools to enrich teaching has witnessed a marked increase. Such tools, especially if they are interactive, easily accessible (e.g., with limited or no installation overhead) and well-designed, can significantly enhance learning.
Nevertheless, the development of interactive visualizations requires significant effort and experience.
The Open Software Services for Classrooms and Research project (OSSCAR, \url{https://www.osscar.org})~\cite{Du2023OSSCARScience} was created to respond to the increasing demand for systematized and
interactive web applications that can be used in both teaching and research contexts, minimizing the development burden for researchers and teachers and maximizing ease of use for students. As described in detail in Ref.~\citenum{Du2023OSSCARScience}, OSSCAR pursues this objective along three main avenues: (1) development of bespoke, interactive Jupyter~\cite{kluyver2016jupyter, JupyterURL}
widgets and notebook extensions, (2) creation of tailored educational content,
(3) detailed documentation. The advantages for end users are manifold. For teachers, there is a vast
simplification of the work required to develop impactful lessons for their classes. For students, the tailor-made lessons facilitate a more effective learning experience. For researchers, leveraging these tools in their day-to-day work makes it possible to streamline a range of tasks.
This paper complements and extends the work presented in Ref.~\citenum{Du2023OSSCARScience} by providing a detailed discussion of the software design decisions that went into creating the widgets that deliver interactive capabilities in OSSCAR. Since the work of Ref.~\citenum{Du2023OSSCARScience}, the Jupyter ecosystem has evolved significantly, including  new major (and backward-incompatible) versions of JupyterLab~\cite{JupyterLabURL}  being released, and the development of new software libraries to facilitate widget development. We have therefore reimplemented, almost from scratch, our widgets to be compatible with the modern Jupyter ecosystem and to use these new libraries for widget development, with the aim of increasing their robustness and ease of development.
Based on this experience, our objective is to provide here also helpful insights for developers of analogous customized widgets, offering a reference for their own development. Therefore, as opposed to focusing on how bespoke widgets can be used to enhance interactive lessons and streamline research workflows, here we explicitly discuss the technical aspects of creating new widgets, and the best practices to ensure their long-term maintainability.
We illustrate our efforts by detailing the role played, within the OSSCAR ecosystem, by our tailored Jupyter widgets (the graphical components that allow for user interaction within Jupyter notebooks) and extensions (the software components that extend the functionality of the JupyterLab interface, that is---as of today---the recommended web-based development environment for Jupyter notebooks). In doing so, we elucidate their role in improving the ease-of-use of the applications in which they have been implemented, and we highlight their
potential for users. The widgets and extensions that we describe in this paper are  not only valuable assets for students, teachers, and researchers in the targeted domains, but they also constitute a framework for streamlined development
of further widgets, aiming at facilitating usage and uptake.

The development of our own widgets was driven by the need of developing OSSCAR notebooks aimed at enhancing lessons with a great degree of interactive visualizations, a goal not easily attainable with the standard available toolbox of widgets, e.g., via the ipywidgets library \cite{ipywidgets}. 
We targeted selected applications, chosen to cover the requirements of a large fraction of the computational physics, chemistry, and materials science communities.
Importantly, considering the current popularity of Jupyter, we approached the development of our toolbox with the goal of creating a general framework for producing user-friendly widgets and notebook extensions, facilitating the notebook development process without exiting the convenient Jupyter ecosystem. 
Our library can easily be further populated, or simply taken as an example and adapted by other teachers or researchers who wish to develop custom web-based tools.

In the following, we first introduce a prototypical
example in which a custom widget is employed within a sample lesson.
We highlight the advantages gained by adopting such a bespoke widget: the improved effectiveness of information transfer to students new to the topic, or to researchers using online tools to assist them
in their work. We then briefly detail the technology underlying
custom widgets. 
We emphasize that while we want to provide a sufficiently in-depth discussion to clarify the considerations that go into creating custom widgets, our goal is not to provide here a user manual or tutorial, which would become quickly obsolete and for which ample resources are available online. We therefore provide only the minimal amount of information necessary to enable readers to better understand the benefits of custom widgets, when it is appropriate to develop new ones rather than using existing components, and the main steps to follow when such customized widgets need to be created. Finally, we present several other OSSCAR widgets and extensions that we developed, each
showcasing unique features which benefit users in several subdomains of computational physics and chemistry. All material presented here is released as open source.

\section{\label{sec:firstexample}
\texttt{widget-bzvisualizer}: A Jupyter widget to visualize the three-dimensional shape
of the first Brillouin zone of periodic crystals}

\begin{figure*}[tbp]
   \centering\includegraphics[width=13cm]{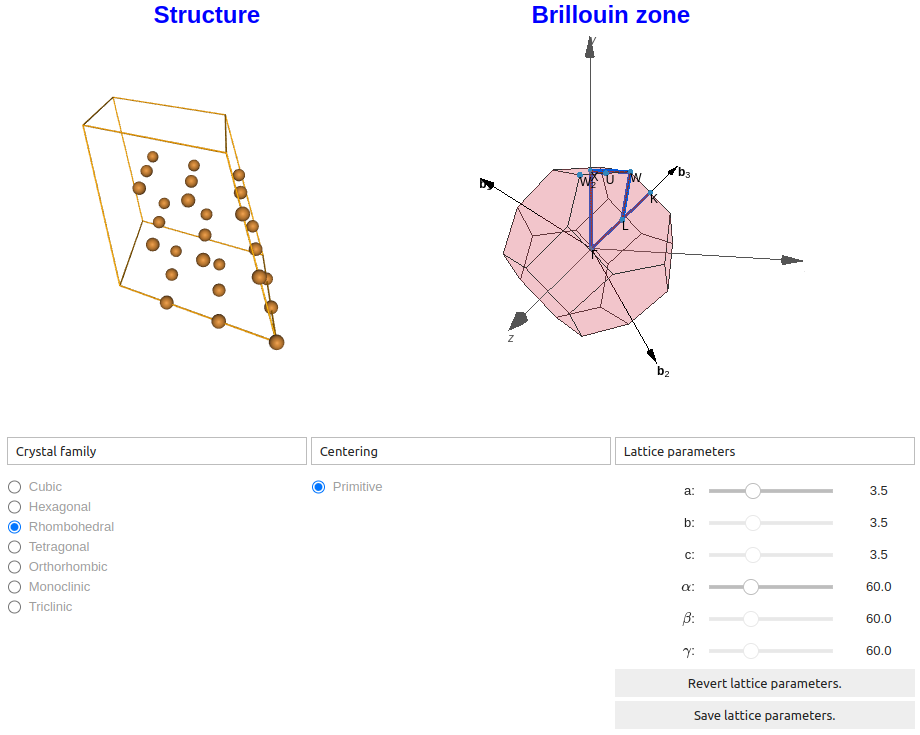}
    \caption{\label{fig:osscar-brillouin-zone}
        Screenshot of a notebook using our custom \texttt{widget-bzvisualizer} widget (top right) to display the 
        first Brillouin zone of a crystal structure. The notebook allows users to compare the appearance of the lattice in reciprocal space (Brillouin zone, right) and in real space (atoms and periodic cell, left, displayed \textit{via} the NGLView widget \cite{NguyenNGLview-interactiveNotebooks,NGLviewWidget}). This 
        notebook is serving as an interactive web application available online at \url{https://www.osscar.org/courses/index.html} (Section ``Band Theory of Crystals'', notebook ``Reciprocal Space and Brillouin Zone'').}
\end{figure*}

In this section, we discuss how we included one of our bespoke widgets into a
notebook, illustrating the concept of reciprocal space in solid-state physics (the notebook can be accessed under the section ``Reciprocal Space and Brillouin Zone'' at \url{https://www.osscar.org/courses/band_theory.html}). This notion, and the related idea of the Brillouin zone (BZ)~\cite{ashcroft2022solid,GrossoBook2013}, are
essential concepts when studying properties of crystalline materials (in which the atoms are arranged on a regular lattice in real space) such as electronic band structures, vibrations and phonons, or magnetic properties. For an in-depth discussion, we refer to any introductory text on solid-state physics, e.g. Ref.~\citenum{ashcroft2022solid}.
The fundamental goal for this notebook is
to convey the dual nature of real and reciprocal space for the description of the system. For
instance, the concept that increasing the size of a crystalline lattice in real
space induces the inverse behavior in reciprocal space. The
real-space and reciprocal-space lattices are potentially complex three-dimensional (3D) objects,
whose construction and plotting is non-trivial. While libraries such as NGLView, see below~\cite{NguyenNGLview-interactiveNotebooks, rose2015ngl, NGLviewMain}, exist to plot crystal lattices in real space, plotting the Brillouin zone in reciprocal space is beyond standard plotting tools such as matplotlib
\cite{Hunter2007Matplotlib:Environment}, unless very lengthy code is implemented.
To most effectively
transfer these notions to students, however, it is useful to have a
notebook-based lesson with side-by-side plotting of the
3D real-space and reciprocal-space lattices. Coupling these
graphical components with a set of interactive elements (such as sliders, radio buttons,
and buttons) enables users to manipulate the parameters of the crystal and obtain immediate visual feedback on the changes. This directly translates relatively abstract concepts and mathematical relationships into interactive 3D visualizations that facilitate understanding. Fig.~\ref{fig:osscar-brillouin-zone} illustrates the key features of the notebook that we developed for these visualizations. The top left element of Fig.~\ref{fig:osscar-brillouin-zone} shows the crystal lattice structure in real space, using an available open-source widget called NGLView\cite{NguyenNGLview-interactiveNotebooks}, one of the few existing visualizers of molecular or solid structures that can be integrated directly into a Jupyter notebook (another popular one is py3DMol\cite{3dmol}). Users can visualize complex materials from input files in a variety of widespread formats (such as XYZ, CIF, and XSF), and interact with them using the mouse to rotate or zoom the displayed structure. 
NGLView constitutes an excellent example of a custom widget that enables many users and researchers to visualize common data types (in this case, crystal structures and molecules), and for this reason we adopt it in our notebooks.
However, no such widget existed for the visualization of the BZ. 
We filled this gap by developing \texttt{widget-bzvisualizer}, which we then used to produce the visual element on the top right of Fig.~\ref{fig:osscar-brillouin-zone}, where the BZ is shown as a pink polyhedron.
This widget (similarly to NGLView) is fully interactive, supporting
the use of the mouse, trackpad, or touchpad to rotate and zoom the 3D visualization.

In the notebook, we also employ radio buttons, sliders, and buttons from the ipywidgets library
to allow users to choose a crystal family and to tune lattice
parameters. Users can select one of the possible crystal families (cubic, hexagonal, tetragonal, etc.) and its centering; six sliders then allow one to adapt the six lattice parameters
(i.e., the length of the three unit-cell vectors $a$, $b$, and $c$, and the angles between them $\alpha$, $\beta$, and $\gamma$).  
Depending on the chosen crystal structure, only the sliders that are not fixed by the constraints of that structure are enabled (e.g., only $a$ for a cubic system, where $b=c=a$ and $\alpha = \beta = \gamma = 90^\circ$).
A button ``Save lattice parameters'' will save the current state (i.e. the values of $a$, $b$, $c$, $\alpha$, $\beta$, $\gamma$), that can be later recovered with the button ``Revert lattice parameters'' (e.g. if one wants to demonstrate the change of many parameters in a single shot).
Composing all these widgets, we created the complete interactive graphical user interface (GUI) shown in Fig.~\ref{fig:osscar-brillouin-zone}. 

The development of custom widgets like NGLView and \texttt{widget-bzvisualizer} requires non-trivial coding, not only in Python (the programming language in which one would implement the logic of the notebook), but also in JavaScript or TypeScript. For instance, fast 3D visualization in the browser using modern WebGL \cite{webgl} technologies is enabled via use of the Three.js JavaScript library~\cite{threejs_url}.
However, according to our experience, JavaScript and its frameworks and libraries are not commonly known by teachers and, even if they are known, developing such interactive visualizations requires expert knowledge of both Python and JavaScript.
Our main goal in developing widgets is to enable users (teachers or researchers) to use such advanced visualization tools without any knowledge of JavaScript. Rather, widgets can be integrated very simply into a Jupyter notebook with very few lines of code, and allow for interaction via intuitive Python code.

\begin{figure}[tb]
\caption{\label{code:widget-bzvisualizer}
Demonstration of the minimal effort needed to use our Brillouin zone visualizer in any Jupyter notebook. (a) The Python code needed to include the Brillouin zone visualizer. (b) The appearance of the
widget when \texttt{disable\_interact\_overlay} is set to true. (c) The appearance of the
widget when \texttt{disable\_interact\_overlay} is set to false: a gray overlay is used to indicate disabled interaction.}
\begin{subfigure}{\linewidth}
\caption{Minimal example of code needed to use the Brillouin zone widget. The code is described in detail in the main text.}
\begin{lstlisting}[language=Python]
from widget_bzvisualizer import BZVisualizer

a = 4.04
cell = [[0., a/2, a/2], [a/2, 0., a/2], [a/2, a/2, 0.]]

w = BZVisualizer(
    cell = cell,
    rel_coords = [[0., 0., 0.]],
    atom_numbers = [13],
    disable_interact_overlay = False
)
display(w)
\end{lstlisting}
\end{subfigure}
\begin{subfigure}{6cm}
    \caption{Widget with \texttt{disable\_interact\_overlay = True}}
    \centering\includegraphics[width=1.0\linewidth]{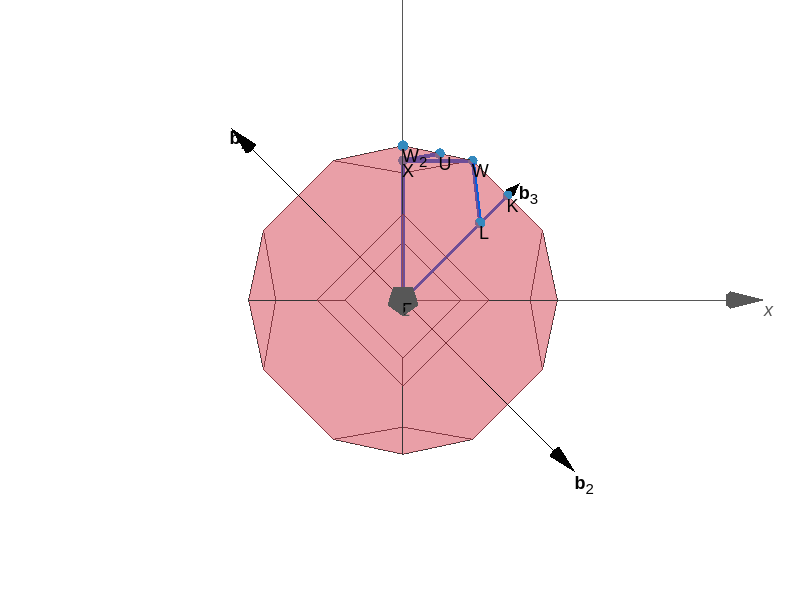}
\end{subfigure}
\begin{subfigure}{6cm}
    \caption{Widget with \texttt{disable\_interact\_overlay = False}}
    \centering\includegraphics[width=1.0\linewidth]{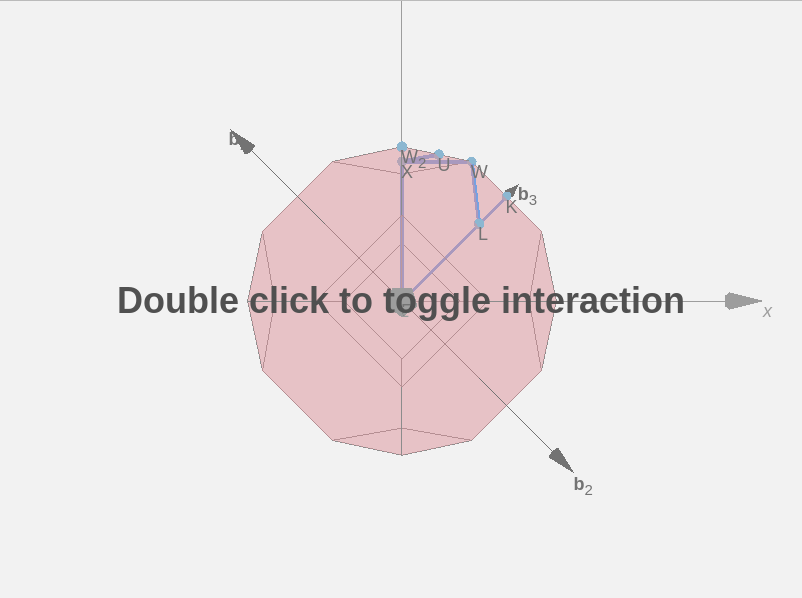}
\end{subfigure}
\end{figure}

To illustrate this point, in Fig. \ref{code:widget-bzvisualizer} we present the usage of \texttt{widget-bzvisualizer} within a Jupyter notebook. 
Apart from importing the required library (line 1), the user only needs to define the real-space cell (a $3\times 3$ array, lines 3-4).
The visualizer is then created by simply passing the lattice and a few additional parameters (that we briefly comment on below) to the \texttt{BZVisualizer} Python class (line 6-11), and then displaying it (line 12).
The library then takes care of computing the reciprocal lattice and the corresponding BZ and rendering it. Furthermore, it attaches standard labels to high-symmetry points, using internally the Python library seekpath~\cite{Hinuma2017BandCrystallography}.
The additional recommended parameters passed to the \texttt{BZVisualizer} include the list of relative atom coordinates in the unit cell (parameter \texttt{rel\_coords} on line 8) and the corresponding list of chemical numbers $Z$ (parameter \texttt{atom\_numbers} on line 9).
In the example, we specify a single aluminum ($Z=13$) atom at the origin (\texttt{[0., 0., 0.]}).
As an advanced note for experts on crystal structures, we emphasize that while the shape of the Brillouin zone is not affected by the number and position of atoms in the real-space unit cell, the space-group symmetry typically depends on the atomic positions.
This is reflected in potentially different labels of the high-symmetry points (also visible in Fig.~\ref{code:widget-bzvisualizer}, that are shown automatically in the visualizer following the conventions of Ref.~\citenum{Hinuma2017BandCrystallography}); for a more detailed discussion, we refer interested readers to Ref.~\cite{Hinuma2017BandCrystallography}.

Additional parameters can be passed to the \texttt{BZVisualizer} to further customize its appearance, such as explicitly specifying the height and width of the widget on the screen.
We further investigated user experience and usability and realized that, since scrolling when the mouse is located inside the widget causes zooming of the visualization, a user may unexpectedly start zooming when their intention is to scroll down the web page.
We therefore also provide an additional boolean parameter \texttt{disable\_interact\_overlay} that can be optionally passed in (line 10) to control whether an overlay appears, and that disables interactivity (as a consequence, scrolling over the widget just scrolls the whole page, rather than zooming in/out). The widget appearance for the two possible values of \texttt{disable\_interact\_overlay} are compared in panels (b) and (c) of Fig.~\ref{code:widget-bzvisualizer}.
When the widget is disabled, a double click re-enables interactivity. Moreover, we highlight that parameters passed to \texttt{BZVisualizer} can be dynamically changed from the Python code after the visualizer is displayed, thus providing an even larger degree of interaction. E.g., one can modify the \texttt{cell}, and this will trigger a recalculation of the Brillouin zone with changes instantly reflected in the visualizer.
Similarly, one can provide users with an additional button to enable or disable the interaction, to address the usability issue discussed above.

\section{\label{sec:developing-widget} The technology behind our Jupyter widgets}

In this section we briefly outline, in general terms, the software components that are common to custom widgets. As mentioned in the Introduction, we do not aim to provide a full documentation or tutorial. Rather, we simply highlight the components that need to be implemented when a widget is developed in order to achieve effective interactivity, and provide readers with a list of references which are useful for this development process. Furthermore, we summarize the recent approaches and libraries that we have adopted in our widget development to facilitate maintainability and accessibility.

\begin{figure}[tb]
    \centering\includegraphics[width=13cm]{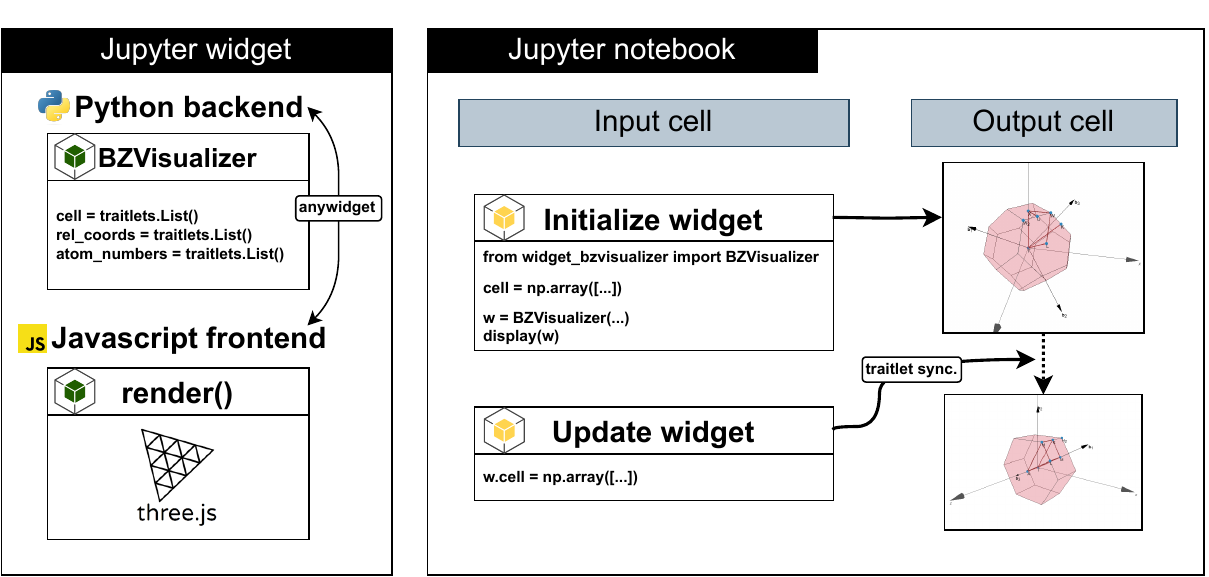}
     \caption{\label{fig:widget-scheme}
     A schematic representation of the code architecture of Jupyter widgets, with the specific example of our BZ visualizer widget (left panel); and the usage of the widget in a Jupyter notebook (right panel).
     The central part of the Python backend is the set of Traitlet attributes, which allow for data synchronization with the JavaScript frontend.
     The \texttt{anywidget} library provides a simple and clean way to link the Traitlets to the JavaScript frontend, which supports any custom visualizations (in this case, for instance, the three.js Javascript library is used to render 3D objects in the browser).
     Users can initialize the Jupyter widget in a Jupyter notebook by writing Python code in a Jupyter input cell. The corresponding visualizations are displayed in the corresponding output cell. The user can then update the widget by modifying the Traitlet attributes (in the same or in a different cell), and the effects are directly reflected in the widget visualization.
     }
 \end{figure}

The code underlying a typical widget consists of two main parts, namely the Python backend and the JavaScript frontend. The Python backend defines the widget class, its attributes, and specifies how the user is expected to interact with the widget from Python code. It is connected to the Jupyter Python kernel to facilitate linking with the notebook. 
The JavaScript frontend, instead, is responsible for constructing the
graphical representation of the widget that is then visualized in the Jupyter notebook. See Fig.~\ref{fig:widget-scheme} for a schematic representation of the two parts.

In the Python backend component, the central technology is the Traitlets package \cite{traitlets}. This library facilitates synchronization of data between frontend graphical elements and the widget attributes defined in the Python backend. To be more specific, the relevant widget properties (i.e., the attributes of the widget's Python class, such as the cell of the Brillouin zone visualizer, the value of the \texttt{disable\_interact\_overlay} flag, \ldots) are stored as a Traitlet variable (e.g. a Traitlet \texttt{List} class to wrap a Python list). These Traitlet variables are made visible and synchronized with the JavaScript side whenever they are modified in the backend, so that the frontend can update the widget accordingly. This is the underlying mechanism that enables, for instance, to recompute and replot the Brillouin zone when the widget \texttt{cell} attribute is changed, as discussed earlier. This reaction to data changes is shown schematically in the right panel of Fig.~\ref{fig:widget-scheme}.

The aforementioned software components form the foundation of any Jupyter widget. Tools are available that provide templates to develop these components and of the syntax needed to link the Python backend with the JavaScript frontend. In earlier versions of our widgets, we directly used the templating tools provided by the ipywidgets library  ~\cite{ipywidgets,ipywidgets_custom_widgets}. Following experimentation with the development of several widgets using various approaches, we found that the technology that provides the most streamlined development process and long-term maintainability is the \texttt{anywidget} library \cite{Manz_anywidget}. \texttt{anywidget} provides a clean and straightforward interface of linking the widget backend and frontend, abstracting away any complicated configuration and boilerplate code. Additionally, it provides templates to quickly get the new widget up and running. Alternative technologies do exist, such as \texttt{widget-cookiecutter} \cite{cookiecutter_template}, that was used to construct the initial versions of our widgets. However, in our experience \texttt{anywidget} provides a simpler and more robust platform for widget development, including a much simpler migration path when supporting new JupyterLab versions. Moreover, owing to the fact that the standard Javascript version of \texttt{widget-cookiecutter} is now deprecated, \texttt{anywidget} benefits from a greater level of support.  
We stress that in general, when developing widgets, one should try to reuse as much as possible existing Python and JavaScript libraries (for the backend and frontend, respectively).
For example, in the case of the BZ visualizer, we employed the existing \texttt{seekpath} Python package~\cite{Hinuma2017BandCrystallography} specifically designed to automate the calculation of the coordinates of the BZ of a crystal. 
Similarly, to facilitate the reuse of the frontend components also beyond Jupyter notebooks, we follow a development pattern in which we release the
JavaScript components of several of our custom widgets as NPM packages \cite{npm_website} (NPM is the most widely used package manager and registry of JavaScript libraries). For example, we released the JavaScript code used to construct and render the BZ widget as the NPM package \url{https://www.npmjs.com/package/brillouinzone-visualizer}. 
Having a separate NPM package makes it possible to reuse the visualization components in other web applications, even when they are not using Jupyter (as it is the case, e.g., for the Seekpath webpage available at \url{https://www.materialscloud.org/seekpath}, that uses the same underlying NPM library for BZ visualization directly in JavaScript).
At the same time, separating the JavaScript code in a different package also helps to keep the amount of code in the widget repository relatively small, and limited to the Jupyter-specific parts (interaction between backend and frontend, as described earlier in this section). This has the additional benefit of facilitating maintenance and integration with future upgrades to newer versions of Jupyter or JupyterLab.

 As of the time of writing, we have thoroughly tested the functionality of all widgets (and OSSCAR applications which employ them) in the most recent versions of JupyterLab (v4) and Notebook (v7), and we support these versions. We do not guarantee, though, that all widgets work in outdated versions of JupyterLab or Jupyter Notebook. The justification for settling on these software dependencies is that, due to the constant evolution of the various technologies on which our software relies upon, incompatibilities between dependencies become inevitable. Therefore, to ensure a reliable experience for end-users of our widgets, we opted to test and support only the most current stable versions of these technologies, and enforce a well-defined set of requirements to guarantee long-term stability of our widgets and notebooks.

\section{\label{sec:widgets}Additional custom widgets}
In this section, we describe two additional custom widgets that we developed for use in Jupyter notebooks. 

They both address very common and widespread use cases, for which we could not find existing tailored widgets: 1) displaying an interactive periodic table that allows to select multiple elements, and 2) visualizing the band structure of a crystal together with the corresponding (partial or full) density of states (DOS).

\subsection{\texttt{widget-periodictable}: A Jupyter widget to display a selectable interactive periodic table}
The periodic table of elements is a fundamental tool, used at some stage in any physics, chemistry or materials science class, e.g. to rationalize patterns amongst different element 
properties (electronegativities, ionization energy, metallic character, etc.). The periodic table is also constantly used in research contexts. For example, various databases of chemical compounds such as Aflowlib \cite{curtarolo2012aflowlib}, Materials Cloud \cite{Talirz2020MaterialsScience}, Materials Project \cite{Jain2013Commentary:Innovation, materials_cloud_website}, and OQMD \cite{saal2013materials, kirklin2015open} employ an interactive periodic table as an element 
filter for fast and interactive searches of catalogued compounds.
However, to the best of our knowledge, there are no existing Python packages to deploy an interactive periodic table inside a Jupyter notebook. To fill this gap, we developed \texttt{widget-periodictable}.

\begin{figure}[tb]
    \caption{\label{fig:widget-periodictable}
    A demonstration of a code snippet to show our interactive \texttt{widget-periodictable} in a Jupyter notebook. (a) The Python code to import and visualize the widget and
    initialize the periodic table. In this example, we allow the user to put any element in one of two possible states (in addition to the ``unselected'' default state); these states are numbered 0 and 1, and displayed with a green and red color, respectively. Elements Si and Ge are preselected in the first state (0), while C is preselected in the second state (1); this could represent, for instance, the example where a user wants to search for all elements in a database including silicon and germanium, but not including carbon. Moreover, five elements (hydrogen and four noble gases) are disabled for user interaction (and thus shown in gray). (b) The appearance of the interactive periodic table obtained by running the Python code in (a).}
    \begin{subfigure}{\linewidth}
    \begin{lstlisting}[language=Python]
    from widget_periodictable import PTableWidget
    
    w = PTableWidget(states = 2, 
                     selected_elements = {"Si": 0, "Ge": 0, "C": 1}, 
                     selected_colors = ['green', 'red'], 
                     disabled_elements = ['H', 'Ne', 'Ar', 'Kr', 'Xe'],
                     unselected_color='pink', border_color = 'black')
    display(w)\end{lstlisting}
    \caption{Python code.}
    \end{subfigure}
    
    \begin{subfigure}{13cm}
       \centering\includegraphics[width=0.8\linewidth]{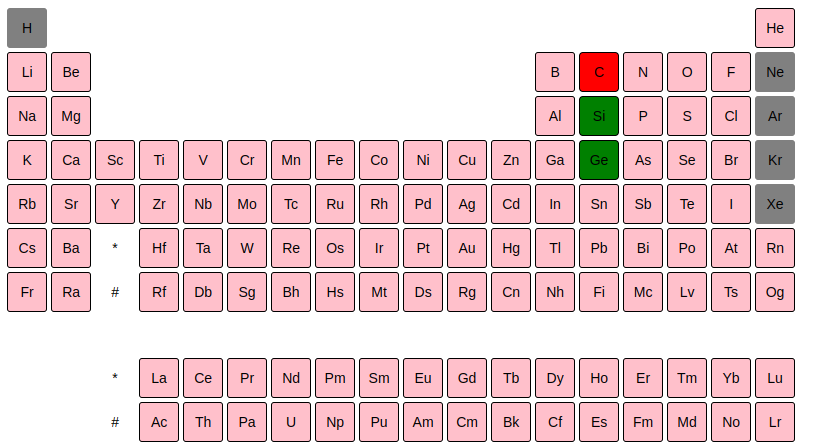}
       \caption{Widget appearance.}
    \end{subfigure}
    \end{figure}

Fig. \ref{fig:widget-periodictable} shows how the widget
can be used.
We designed it with functionality to allow users to group elements into 
different states, identified by custom colors.
As a concrete example of how this functionality may be used, a user of the widget might wish to query a database of materials for specific compounds containing elements X and Y, but specifically not Z. Configuring the widget to support two states, setting elements X and Y to state 0 (``include in search'', e.g. colored in green) and element Z to state 1 (``exclude from search'', e.g. colored in red) would allow them to intuitively achieve this.
Toggling between the unselected state and any of the selected states is achieved by clicking on the element. 
The selected elements and their states can be retrieved at any moment directly from Python code to enable further interaction or, conversely, elements can be set in a specific state from Python code.
We also provide functionality to disable specific elements, preventing users from selecting them or toggling their states (useful, for instance, if the database being queried does not support those elements). Disabled elements are displayed in a customizable color, gray by default, as it is the case for the elements H, Ne, Ar, Kr and Xe in Fig.~\ref{fig:widget-periodictable}(b). 

\begin{figure}[tbp]
   \centering\includegraphics[width=0.8\linewidth]{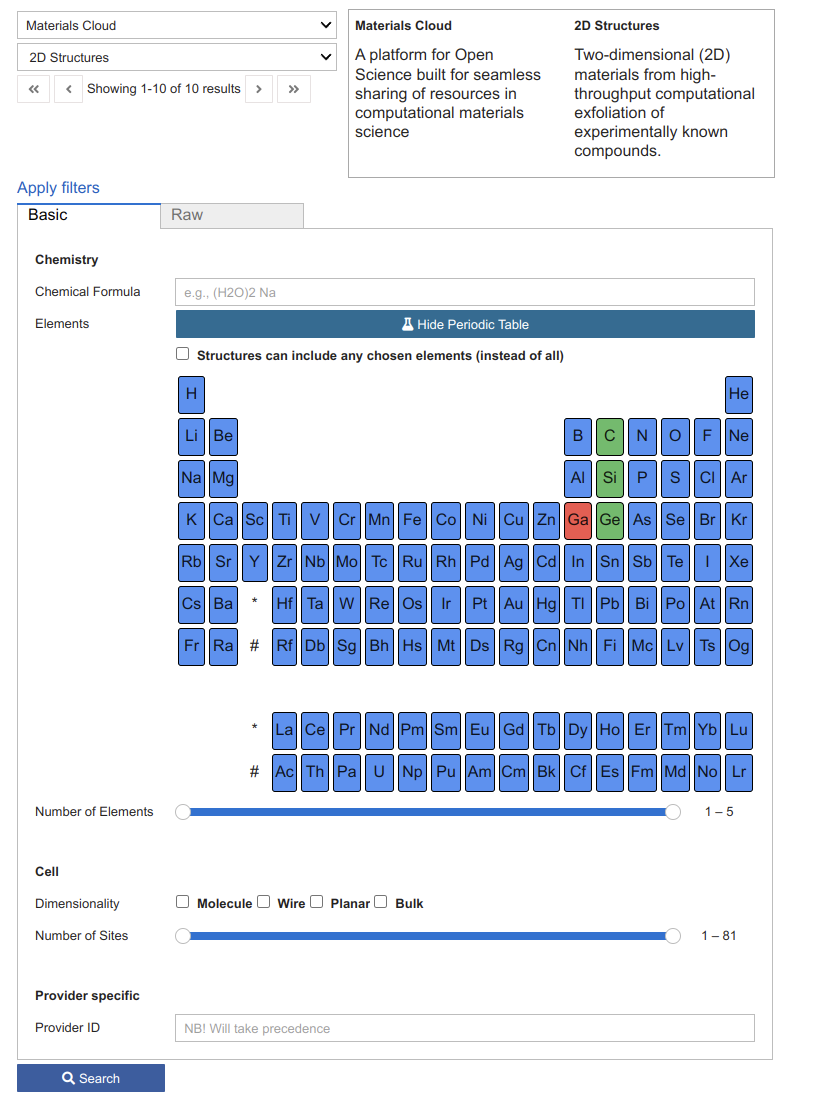}
    \caption{\label{fig:optimade}
    The \texttt{widget-periodictable} embedded as an interactive element
    filter for the OPTIMADE client web application \cite{optimade_client}. 
    A single click on an element selects it for inclusion in the search, turning it green, and requiring that any search result include that element; an additional click selects it for exclusion (red state), while a third click unselects the element, turning it back to a blue background.
    }
\end{figure}

This type of searching and filtering functionality may also find use in educational contexts. For instance, students in a course might be asked to select elements within a given group or period in the table, while dynamically displaying or plotting the properties of selected elements in a different part of the web page. This would offer them the opportunity of visually identifying chemical trends, offering an effective learning strategy to familiarize themselves with fundamental associations between chemical elements and their properties.
A notable application of the  \texttt{widget-periodictable} in a research context is its use as an element selector to search and filter materials inside the
OPTIMADE (Open Databases Integration for Materials Design)
client app \cite{optimade_client}. This client app is a Jupyter-based GUI to facilitate searches of crystal structures in the materials databases that expose the common OPTIMADE query API \cite{optimade_website, Andersen2021OPTIMADEData,OPTIMADE2024} hosted on Materials Cloud~\cite{Talirz2020MaterialsScience} at \url{https://www.materialscloud.org/optimadeclient}.
Fig. \ref{fig:optimade} shows how the periodic table widget appears within the OPTIMADE client where, similarly to the example given earlier, two states (shown as green and red respectively) correspond to whether that element should be included or excluded in a search.
Inclusion of the periodic table widget greatly enhances accessibility of this GUI by facilitating easy filtering of elements.

\subsection{\texttt{widget-bandsplot}: A Jupyter widget to plot and visualize electronic band structures and density of states}

\begin{figure}[tbp]
\caption{\label{fig:widget-bandsplot}
A demonstration of a
code snippet to show our interactive \texttt{widget-bandsplot} in a Jupyter notebook. (a) The Python code to import and
visualize the widget, and show band structure and DOS data that were already computed (stored in the variables \texttt{band\_data1}, \texttt{band\_data2} and \texttt{dos\_data}). Two 
band-structure data sets are shown together (\texttt{band\_data1} and \texttt{band\_data2}) to allow for comparison. They are shown as
red and yellow curves, respectively, as specified by the parameter \texttt{bands\_color}. Additional parameters are specified (e.g. whether to show a horizontal line at the Fermi level, or the energy range to display). (b) The appearance 
of the deployed \texttt{widget-bandsplot} obtained running the Python code in (a), in the case of the bands and DOS of Silicon.}
\begin{subfigure}{\linewidth}
\begin{lstlisting}[language=Python]
from widget_bandsplot import BandsPlotWidget

w = BandsPlotWidget(
    bands=[band_data1, band_data2],
    dos=dos_data,
    energy_range = [-10.0, 10.0],
    bands_color=['orangered', 'skyblue']
)
display(w)
\end{lstlisting}
\caption{Python code.}
\end{subfigure}

\begin{subfigure}{1.0\linewidth}
   \centering\includegraphics[width=1.0\linewidth]{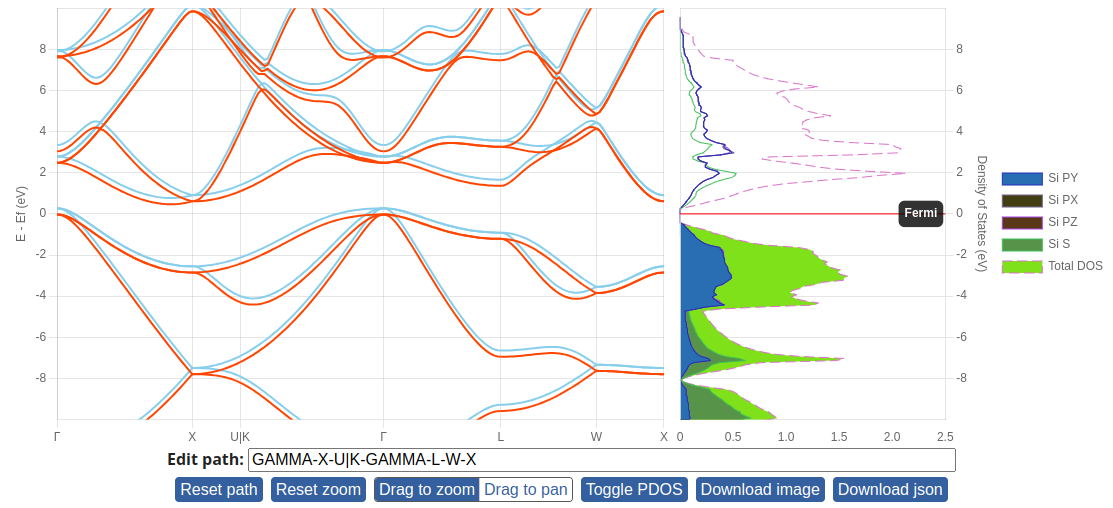}
   \caption{Widget appearance.}
\end{subfigure}
\end{figure}

The electronic band structure and density of states (DOS) are two central properties of study in solid-state systems \cite{ashcroft2022solid, kittel2018introduction}. They are used to describe occupied and unoccupied energy levels in a solid, and therefore provide insight into its electronic and optical properties, for instance. 
Band structure and DOS are typically obtained from electronic structure calculations (e.g., based on density functional theory, tight binding or other theoretical frameworks), but their (simultaneous) visualization is non-trivial.
The energy levels in a band structure are given as a function of the 3D electronic wavevector $\bm{k}$.
Since it is not possible to plot a function of a 3D variable, one typically plots the band structure of a solid following a path connecting high-symmetry points in its BZ, resulting in figures like that shown in the left-hand side of Fig.~\ref{fig:widget-bandsplot}(b). 
On the other hand, the DOS is a
density distribution of the allowed energy levels. Often, both band structure and DOS figures are plotted side by side, sharing the vertical energy axis as shown in Fig.~\ref{fig:widget-bandsplot}(b), where the DOS is represented on the right-hand side.

Researchers generally need to use software or scripts to extract and analyze the raw data and convert it into band structure and DOS
plots. Despite the existence of various cloud-based databases (e.g., AFLOW \cite{curtarolo2012aflowlib}, Materials Cloud
\cite{Talirz2020MaterialsScience}, Materials Project \cite{Jain2013Commentary:Innovation} and OQMD \cite{saal2013materials, kirklin2015open}), which offer high quality visualizations of these quantities, it is not trivial to produce such plots in very few lines of code within Jupyter notebooks.
We therefore developed a Jupyter widget called \texttt{widget-bandsplot}, to generate interactive figures showing the electronic band structure and density of states. 
The widget uses the Chart.js JavaScript library \cite{chart_js} as the underlying frontend plotting library.
Fig.~\ref{fig:widget-bandsplot}(b) illustrates how to use the widget in a scenario where both the band structure and the DOS are plotted simultaneously (we note that the widget also allows one to plot the two separately), together with the final appearance of the widget.
For the band structure component of the widget, there is support for visualizing with different colors the bands with different spin polarization (i.e., spin-up and spin-down bands for simulations without spin--orbit coupling and with collinear spin treatment). Moreover, users can plot and compare multiple band structures within the same figure (e.g., to compare the band structure of different materials, or for the same material but using different theoretical frameworks). 
As a simplified means of specifying and modifying the k-points path to use for the plot, we also provide a textbox widget below the main widget pane (see Fig.~\ref{fig:widget-bandsplot}). Users can edit the text within this widget to effortlessly select, swap and reorganize the list of high-symmetry paths used in the band plot. 
The DOS component of the widget, on the other hand, has the capacity to display not only the total DOS, but also the projected DOS (PDOS) within the same figure. This helps users discern the contribution of individual electronic states (from different atoms or orbitals) to the total DOS. The spin polarization associated with a given PDOS is shown by having positive values for spin-up states and negative values for spin-down states.

An example of the data format is provided explicitly in the README file of the GitHub repository of the widget \cite{bandsplot_repo}.
Both figures are interactive: users can employ their mouse to accurately pick out the coordinates of specific data points, to toggle the visibility of the projected density of states, and more. 
In addition, when bands and DOS are plotted together, the vertical energy axis is shared and synchronized (i.e., zooming on one plot will also zoom on the other) to facilitate the investigation of the contribution of various bands to the DOS. 

To use the widget, it is necessary to first load the input files containing the band structure data and the DOS data as Python dictionaries (possibly serialized to file in the JSON format, for instance).
In order to simultaneously plot the bands of multiple materials, or resulting from different calculations, users can simply provide as input the different band structure dictionaries. 
For example, the band structure data of Si obtained from two separate calculations are given as arguments to the attribute ``bands'' of the \texttt{BandsPlotWidget} object as shown in 
Fig.~\ref{fig:widget-bandsplot}(a). The corresponding bands are shown simultaneously with different colors, see
Fig.~\ref{fig:widget-bandsplot}(b).
The color of bands and of DOS lines can be customized either from the input file or specifying parameters to the \texttt{BandsPlotWidget}.

The \texttt{widget-bandsplot} constitutes a versatile visualization tool which can be used in a variety of contexts.
In addition to being used to analyze results of \textit{ab initio} calculations in Jupyter notebooks, as shown in Fig.~\ref{fig:widget-bandsplot}, the underlying Javascript library is also used in the Materials Cloud platform to visualize electronic properties in curated databases (such as in the Materials Cloud 2D crystals database \cite{Mounet2018,Campi2023,mc2d}). 
Furthermore, this custom widget has also been transferred, in the spirit of the OSSCAR project, from the research context for which it was originally designed, to an educational context. The \texttt{widget-bandsplot} is employed in several of our interactive lessons as a vehicle to teach materials-science students about the fundamental concepts of electronic band structure and density of states. For example, in the OSSCAR Quantum Mechanics notebooks \cite{osscar_qe_notebooks}, a notebook is devoted to investigation of the band structure of a free-electron gas in periodic boundary conditions (\url{https://osscar-quantum-mechanics.materialscloud.io/voila/render/band-theory/free_electron.ipynb}). This notebook enables illustration of the effect of band folding and the similarity of free-electron bands to the actual bands of some simple crystals.

\section{\label{sec:examples}Beyond widgets: JupyterLab extensions}
JupyterLab is currently well positioned as the next-generation notebook interface.
It provides a flexible interface for interactive programming and notebook editing, supporting multiple notebook editing and visualization at the same time, direct inspection and visualization of files, and much more.
JupyterLab also supports the development of extensions to customize almost all aspects of its interface. 

The goal of these extensions is distinct from that of widgets. 
Widgets, as we saw in the earlier sections, are components that are added within a notebook: they have a Python interface to interact with them, and when called from the notebook input cells, they generate different types of custom interactive visualizations in the output cell, to display the data being analyzed in the notebook.
Extensions, instead, augment JupyterLab on a global scale, altering how the JupyterLab interface is used and thus supplementing the experience of JupyterLab users. 
Hundreds of JupyterLab extensions have already been
developed and can be easily installed, similarly to widgets, by using the \texttt{pip} \cite{pip} package manager to fetch them from the PyPi repository \cite{pypi}.
Extensions can, for instance, change the graphical appearance (or theme) or the JupyterLab interface, add new file viewers with rich renderers for specific file types, or add new buttons, command palettes or menu items.

In this section, we discuss one example of a JupyterLab extension that we developed to enhance the experience of a broad research community, illustrating the flexibility of the OSSCAR toolbox.

\subsection{\texttt{jupyterlab-mol-visualizer}: A JupyterLab launcher to visualize molecules and molecular orbitals}

Molecular orbitals (MOs) are a useful tool to gather chemical insight from the results of quantum chemistry calculations. 
The numerical representation of molecular
orbitals is usually provided by volumetric data (i.e., a discretized set of values adopted by some quantity, like the electronic density, on a grid of points in 3D). This data is typically displayed as isosurfaces in a 3D visualizer.
The popular quantum chemistry software Gaussian \cite{g16}, for example, introduced the ``cube file'' format to output volumetric data for molecular
orbitals and electron densities. The format is now widely adopted by many other quantum chemistry and solid-state physics codes, such as 
VASP\cite{Kresse:1996,Kresse:1999}, Quantum ESPRESSO~\cite{Giannozzi2009QuantumMaterials}, 
Dalton~\cite{Aidas2014TheSystem}, Orca~\cite{Neese2020ThePackage}, and many more.
Visualization of these cube files is therefore a daily task carried out by computational physics and chemistry researchers.

Most standalone molecular and crystal visualizers can display cube files in their own interface. However, this happens outside of the JupyterLab interface, so that users analyzing simulation data in JupyterLab must change interface to perform the visualization.
Alternatively, it is possible to visualize 3D volumetric data as widgets within Jupyter notebooks using, for example, Pythreejs \cite{pythreejs}, Paraview \cite{paraview}, or NGLView \cite{NguyenNGLview-interactiveNotebooks}. Using these tools, however, requires creating a new notebook every time one wants to visualize a cube file, and writing multiple lines of code to generate the visualization.

\begin{figure*}[tb]
   \centering\includegraphics[width=8.5cm]{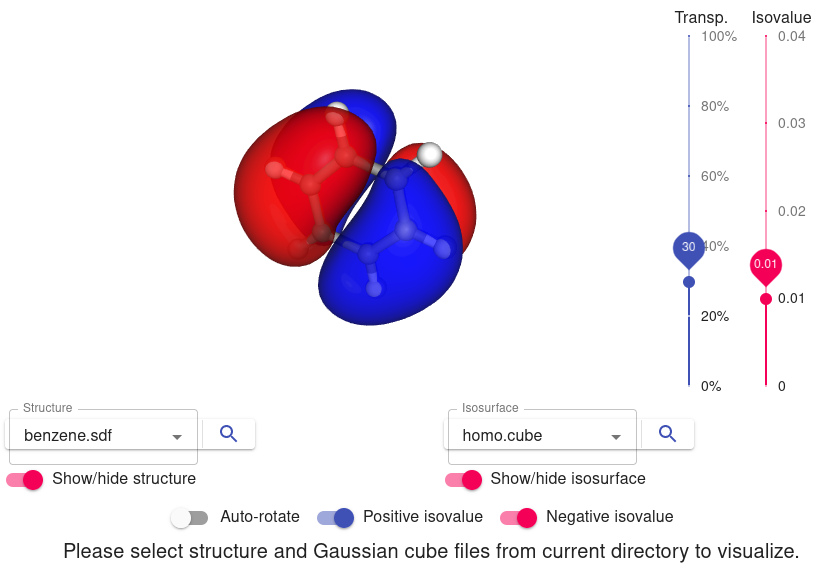}
    \caption{\label{fig:jupyterlab-mol-visualizer}
    The JupyterLab extension \texttt{jupyterlab-mol-visualizer} is demonstrated in this screenshot for the visualization of a benzene molecule and the isosurface of its highest-occupied molecular orbital (HOMO).
    Various tools facilitate the inspection of the 3D model:
    the blue slider on the left tunes the
transparency of the MO isosurface, while the red slider selects the value of the targeted isosurface. The bottom controls allow for the selection of different files for the molecular structure and the isosurface, and to tune various visualization aspects.}
\end{figure*}

To simplify the visualization and data analysis workflow, we thus developed a bespoke JupyterLab launcher extension, called \texttt{jupyterlab-mol-visualizer}, implemented using the NGLView JavaScript library. This extension does not require the user to write any code, nor to leave the JupyterLab interface in order to visualize a cube file.
Instead, user can simply select and open a file with \texttt{.cube} extension (from the current working directory) directly in the JupyterLab interface, obtaining an interactive 3D visualization.
We show an example of the extension in Fig.~\ref{fig:jupyterlab-mol-visualizer}, displaying the highest occupied molecular orbital (HOMO) of benzene. Using the two white dropdown boxes shown in the figure, users can easily change the choice of molecular orbital file, and also select the corresponding structure file (in XYZ, CIF and SDF format) to visualize orbitals together with the atoms of the molecule. The widget also provides additional tools to facilitate inspection of orbitals, as discussed in the caption of Fig.~\ref{fig:jupyterlab-mol-visualizer}.
We note that, in addition to providing an embedded visualizer for the common cube format, our \texttt{jupyterlab-mol-visualizer} extension constitutes a useful template for the development of similar advanced visualization tools.

\subsection{Extensions to create a custom JupyterLab experience}

\begin{figure*}[tb]
   \centering\includegraphics[width=1.0\linewidth]{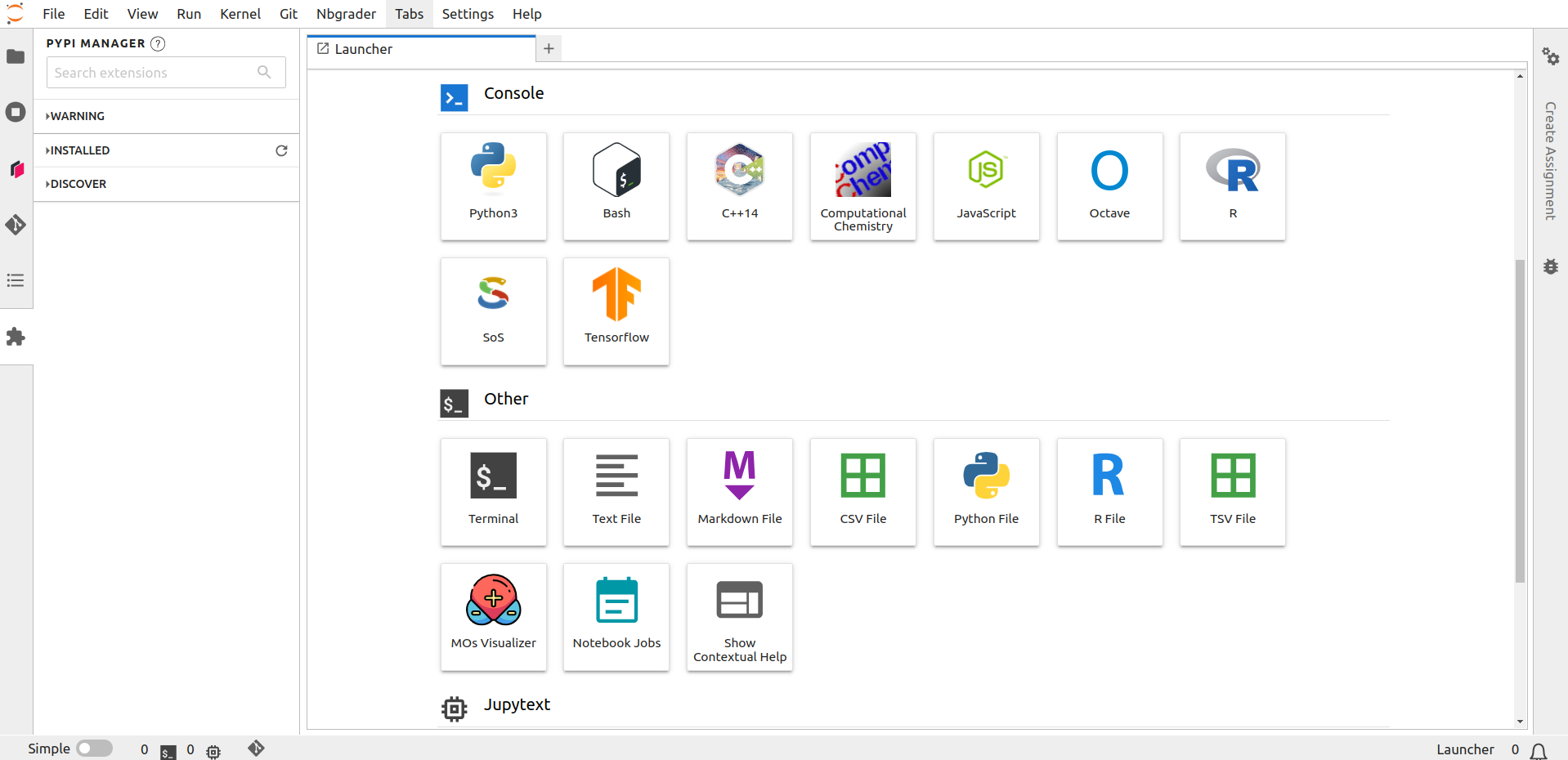}
    \caption{\label{fig:noto-interface} 
    The customized NOTO JupyterLab interface at EPFL. Custom programming environments and 
    tools are provided at the launcher webpage. The \texttt{jupyterlab-mol-visualizer} is preinstalled and
    listed in the ``Other'' section (first button of the last row in the screenshot).}
\end{figure*}

We highlight that the use of custom JupyterLab extensions provides users with a tailored JupyterLab experience; multiple extensions can then be combined to provide a bespoke, ready-to-use programming and 
computing environment for students or researchers of a given discipline. A paradigmatic example of JupyterLab extensions being used in this manner is the NOTO JupyterLab web service offered by the \'Ecole Polytechnique F\'ed\'erale de Lausanne (EPFL) to its students and 
researchers \cite{noto}. As shown in Fig.~\ref{fig:noto-interface}, the
NOTO JupyterLab interface is customized to provide extra tools and environments for education and research in its sidebar and in the launcher (notebooks and
consoles, for example). 
The
\texttt{jupyterlab-mol-visualizer} extension is integrated with NOTO, and it is listed in the ``Other'' section of
the JupyterLab launcher, making its use immediately available to all users of the platform, and in particular to the computational chemistry community of the University. NOTO is based on JupyterHub \cite{jupyterhub}, a technology that has been adopted by an increasing number of institutions around the globe. 
Institutional JupyterHubs thus provide an effective means of bringing extensions like \texttt{jupyterlab-mol-visualizer}, described in the previous section, to a large number of users.

\section{Conclusions}

In this work, we have outlined our efforts to extend the existing Jupyter ecosystem to include
specialized components covering the needs of students, teachers and researchers in the broad communities of computational chemistry, physics, and materials science. 
Specifically, we have developed a set of customized Jupyter widgets and
extensions that facilitate performance of various important visualization tasks within these domains. 
Besides discussing the tools that we have developed, we have detailed the advantages engendered by development of tailored Jupyter widgets, from enabling a level of interactivity and versatility in
notebook-based lessons which significantly enriches the quality of teaching, to opening up the possibility of creating high-quality web applications that expedite usage of common domain-specific workflows encountered in day-to-day research tasks.  
Although the set of custom tools that we provide is limited in number, we believe that with the contribution of these high-quality widgets we have paved the way for a swift extension of the set of tools by interested members of the community, providing effective template examples to facilitate further developments.

\section*{CRediT authorship contribution statement}
\textbf{Dou Du}: Methodology, Software, Validation, Formal analysis, Investigation, Writing - Original Draft, Writing - Review \& Editing, Visualization.
\textbf{Taylor J. Baird}: Methodology, Software, Validation, Formal analysis, Investigation, Writing - Review \& Editing.
\textbf{Kristjan Eimre}: Methodology, Software, Validation, Formal analysis, Investigation, Writing - Review \& Editing, Visualization.
\textbf{Sara Bonella}: Conceptualization, Methodology, Writing - Review \& Editing, Supervision, Project administration, Funding acquisition.
\textbf{Giovanni Pizzi}: Conceptualization, Methodology, Software, Validation, Formal analysis, Resources, Writing - Review \& Editing, Supervision, Project administration, Funding acquisition.

\section*{Declaration of Competing Interest}
The authors declare that they have no known competing financial interests or personal relationships that could have appeared to influence the work reported in this paper.

\section*{Acknowledgements}

We acknowledge financial support from the EPFL Open Science Fund via the OSSCAR
project.  We acknowledge CECAM for dedicated OSSCAR dissemination activities.
We acknowledge the NCCR MARVEL (a National Centre of Competence in Research,
funded by the Swiss National Science Foundation, grant No. 205602)
for the support in the deployment of the
applications on the Materials Cloud (via \texttt{dokku}).  The authors are grateful
to Casper W. Andersen, Michele Ceriotti, C\'ecile Hardebolle, Patrick Jermann, Richard Lee-Davis, Nicola Marzari, Pierre-Olivier Vall\`es and Xing Wang for feedback and useful discussions, to Snehal Kumbhar and Elsa Passaro for contributions to some of the underlying JavaScript code, and to Johannes M. Hermann for support in the upgrade of \texttt{jupyterlab-mol-visulizer} to JupyterLab 4.

\bibliographystyle{elsarticle-num}
\bibliography{library}

\end{document}